\documentclass[11pt]{article}
\usepackage{enumerate}
\usepackage{amsfonts}
\usepackage{amssymb}
\usepackage{amsmath}
\usepackage{amsthm}
\usepackage{latexsym}
\usepackage{hyperref}
\usepackage[all]{xy}
\usepackage{nameref}
\usepackage[left=1.25in, right=1.25in, top=1in, bottom=1in]{geometry}
\usepackage{xspace}
\usepackage{cite}

\theoremstyle{definition}
\newtheorem{definition}{Definition}
\newtheorem*{definition*}{Definition}
\newtheorem{remark}{Remark}
\newtheorem*{example*}{Example}

\newtheorem*{defStretchedPrime}{Definition \ref{def:stretched}$^\prime$}

\theoremstyle{plain}
\newtheorem{theorem}{Theorem}

\newtheorem{open}{Open Question}

\newcommand{\eg}{e.\,g.}

\newcommand{\defeq}{\stackrel{def}{=}}

\newcommand{\F}{\mathbb{F}}

\newcommand{\poly}{\text{poly}\xspace}
\newcommand{\cc}[1]{\mathsf{#1}\xspace}

\let\det\relax
\DeclareMathOperator{\det}{det}

\DeclareMathOperator{\Poly}{Poly}

\DeclareMathOperator{\GL}{GL}

\newcommand{\coeff}[1]{\text{coeff}(#1)}
\newcommand{\Burgisser}{B\"{u}rgisser\xspace}

\title{Towards an algebraic natural proofs barrier \\
via polynomial identity testing}

\author{Joshua A. Grochow\footnote{Department of Computer Science, University of Colorado, Boulder, CO, USA and the Santa Fe Institute, Santa Fe, NM, USA,  \texttt{joshua.grochow@cs.colorado.edu} },
Mrinal Kumar\footnote{Department of Computer Science, Rutgers University, Piscataway, NJ, USA, \texttt{mrinal.kumar@rutgers.edu}}, \\ Michael Saks\footnote{Department of Mathematics, Rutgers University, Piscataway, NJ, USA, \texttt{saks@math.rutgers.edu}},
and
Shubhangi Saraf\footnote{Departments of Computer Science \& Mathematics, Rutgers University, Piscataway, NJ, USA, \texttt{shubhangi.saraf@rutgers.edu}}
}

\date{\today}

\begin{document}

\maketitle

\begin{abstract}
We observe that a certain kind of algebraic proof---which covers essentially all known algebraic circuit lower bounds to date---cannot be used to prove lower bounds against $\cc{VP}$ if and only if what we call succinct hitting sets exist for $\cc{VP}$. This is analogous to the Razborov--Rudich natural proofs barrier in Boolean circuit complexity, in that we rule out a large class of lower bound techniques under a derandomization assumption. We also discuss connections between this algebraic natural proofs barrier, geometric complexity theory,  and (algebraic) proof complexity.
\end{abstract}

\section{Introduction}
The natural proofs barrier \cite{razborovRudich} showed that a large 
class of circuit-based proof techniques could not separate $\cc{P}$ from $\cc{NP}$, 
assuming the existence of pseudo-random generators of a certain 
strength. In light of the recent advances in techniques for lower bounds on 
algebraic circuits \cite{AV, koiranChasm, K12, tavenas, KSSregular, KLSS14, KLSS14b, KSFOCS14, GKKS, sarafSurvey, KayalSurvey14, KSLowArity15, KSic15, KS5, KSTopFanIn, KKPS15, ASSSJacobian, KSsmall16, KSfiner, AFSSV, CKSV16, FKS16, GKKSchasm, KSlocal16, KSfew16, KST16cubic, KST16, KNS16, KMS16, PSS}, it is natural to wonder whether our current algebraic 
techniques could plausibly separate $\cc{VP}$ from $\cc{VNP}$, or whether there is 
some barrier in this setting as well. People often hand-wave about an 
``algebraic natural proofs barrier,'' by analogy to Razborov--Rudich, but it 
has not been clear what this means precisely, and to date no such barrier 
is known in a purely algebraic setting (see below for a discussion of the related work by Aaronson and Drucker \cite{AD, ADblog} in a partially algebraic, partially Boolean setting).

There are several difficulties in coming up with such a barrier in the algebraic context. Razborov and Rudich's notion of natural proof has two key 
ingredients: (1) largeness---the technique works to show that random 
functions are hard---and (2) constructivity---deciding whether a function 
satisfies the hypotheses of the technique can be done efficiently given its 
truth table. These two ingredients in combination allow them to make the 
connection to pseudo-random generators. However, in the algebraic world, all \emph{three} of these 
notions are unclear: What should largeness mean in an algebraic context? 
What should constructivity mean? Is there a good algebraic notion of 
pseudo-random generator?\footnote{Although Agrawal's notion of algebraic PRG \cite{agrawalPRG} is useful in its own setting, it is not clear whether it could be used for an algebraic natural proofs barrier, and in fact connecting our 
formulation to Agrawal's PRGs remains an interesting open problem.} In a mixed Boolean-algebraic setting, Aaronson and Drucker \cite{AD, ADblog} provide satisfactory answers, but in a purely algebraic setting finding a constellation of three answers to these questions that align to give a satisfying algebraic natural proofs barrier has been an open 
question for more than twenty years.

In the purely algebraic setting---algebraic circuits over an arbitrary field---we take largeness to mean Zariski-openness (the complement of the zero set of a set of polynomial equations), and constructivity to mean 
that the property is computable by an algebraic circuit whose size is 
polynomial in the number of coefficients of the function being tested. These 
two properties cover essentially all known algebraic lower bounds to date 
\cite{GrochowGCTUnity, ADblog} (see also \cite[Section~3.9]{SYsurvey}). Rather than connecting these notions directly to 
PRGs, we connect them to a slightly different derandomization problem, 
but one that is natural from the algebraic viewpoint: polynomial identity 
testing (PIT). Here, we suggest that the coefficient vectors of random linear 
projections of the determinant (respectively, a generic algebraic circuit) should 
produce good hitting sets for restricted versions of PIT.  (This is closely related to Aaronson and Drucker's suggestion that they form a pseudo-random family of algebraic functions \cite{AD, ADblog}; see Section~\ref{sec:related} for details.) We observe that if this 
is true, then many proof techniques---including those of the recent 
advances---cannot be used to separate $\cc{VP}_{ws}$ from $\cc{VNP}$. As in the original natural proofs barrier, we thus show that a strong enough derandomization assumption implies that certain techniques cannot prove strong lower bounds. We recently learned that Forbes, Shpilka and Volk came to the same connection with PIT independently, and were able to show some of the derandomization assumptions unconditionally \cite{SV}.

In the final two sections, we comment on how this algebraic natural proofs barrier bears on geometric complexity theory, and how it might be used to prove lower bounds in (algebraic) proof complexity.

\subsection{The idea} \label{sec:idea}
Almost all algebraic circuit lower bounds to date proceed either by the substitution method, or by a ``rank-type'' method, namely: associate to each polynomial $f$ some matrix $M(f)$---\eg, a matrix of partial derivatives or shifted partial derivatives, perhaps exponentially large---show an upper bound on the rank of $M(f)$ for any $f \in \mathcal{C}_{easy}$, and show a lower bound on the rank of $M(f_{hard})$ for some polynomial $f_{hard}$. In all examples to date, the entries of the matrix $M(f)$ are linear functions of the coefficients of $f$; as the rank of $M(f)$ is determined by the vanishing of its minors, we can view this method as an instance of the following more general ``polynomial method.'' For a polynomial $f$, let $\coeff{f}$ denote its coefficient vector. The polynomial method is to find a ``meta-polynomial'' $T$ (called ``test polynomials'' in \cite{GrochowGCTUnity})---whose variables are the coefficients of polynomials $f$---such that $T(\coeff{f})=0$ for all $f \in \mathcal{C}_{easy}$, but $T(\coeff{f_{hard}}) \neq 0$. 

The first step here---as in the original natural proofs barrier---is to consider the circuit complexity of the meta-polynomials $T$ themselves, relative to their number of inputs. The new idea here is to consider which classes of meta-polynomials have $\{\coeff{f} : f \in \mathcal{C}_{easy}\}$ as a hitting set.

Let us consider what this looks like for the rank-type methods mentioned above. Suppose that we are considering polynomial families $f=(f_n)_{n=1,2,3\dotsc}$ in $n^c$ variables of degree $n$. The space of such polynomials has dimension $\binom{n + n^c -1}{n} = 2^{\Theta(n \log n)}$. Since this will be the number of variables of our meta-polynomials (we might call them ``meta-variables''), let us denote it by $N$. The matrices $M(f)$ typically have dimension $\poly(N)$, which is still $2^{\Theta(n \log n)}$. If we are considering whether or not $M(f)$ has rank $\leq r$ or $> r$, then we are considering the (non)vanishing of the $(r+1) \times (r+1)$ minors, which are themselves determinants of size at most $\poly(N) \times \poly(N)$. Therefore, these meta-polynomials lie in the circuit class we denote $\cc{VP}_{ws}(N)$, which is defined just like $\cc{VP}_{ws}$, but where everything---degree, circuit size, etc.---is measured as a function of the number of variables $N$. This circuit complexity upper bound on the meta-polynomials is the algebraic analogue of Razborov and Rudich's constructivity criterion.

Now, suppose we want to prove a lower bound against some class $\mathcal{C}_{easy}$ using such a rank-type argument. If the coefficient vectors of polynomials in $\mathcal{C}_{easy}$ form a hitting set (perhaps infinite) for $\cc{VP}_{ws}(N)$, then no meta-polynomial as in the preceding paragraph can vanish on $\mathcal{C}_{easy}$, precluding such arguments. 
This is the fundamental connection we advance between algebraic circuit lower bounds (by the polynomial method) and polynomial identity testing.

\subsection{Relationship with previous work} \label{sec:related}
Efremenko, Landsberg, Schenck, and Weyman \cite{ELSW1, ELSW2} proved unconditionally that the method of shifted partial derivatives cannot prove a lower bound stronger than $\Omega(n^2)$ on the permanent versus determinant problem. While parts of their methods are not specific to these polynomials, their results \emph{are} specific to the method of shifted partial derivatives. In contrast, our general framework has the potential to rule out proving lower bounds by \emph{any} method where the meta-polynomials are easily computable. While in this paper all our results are conditional, some of them are made unconditional in Forbes--Shpilka--Volk \cite{SV}.

Aaronson and Drucker \cite{AD, ADblog} (see Aaronson's survey \cite[Section~6.5.3]{Asurvey} for an overview) had similar ideas, but ours differ in several respects. One strength of their work compared to ours is that they considered not just algebraic, but mixed Boolean-algebraic settings---that is, considering polynomials over finite fields as Boolean functions of the bitwise description of the field elements---and this allowed them to draw equivalences between the existence of Boolean and (suitably formulated) algebraic pseudo-random functions. In contrast, our work is purely algebraic, and rather than using pseudo-random generators, we use hitting sets for polynomial identity testing. 

The difference between their work and ours which allows us to make the connection with PIT is as follows. They considered a polynomial family $f_n$ to be pseudo-random if it could not be distinguished from a random polynomial family of similar degree by any meta-polynomial computed by small circuits $C_n$, in the sense that $\Pr[f_n(C_n(f_n(x)))=f_n(x)]$ was negligible as a function of $n$ (smaller than $1/n^c$ for any $c$) \cite[Slide~8]{AD}. In order for this to make sense, they considered polynomial families $(f_n)$ over fields of growing size $\F_{p(n)}$ (and the probability is taken uniformly over $x \in \F_{p(n)}^n$). This is quite close to the usual Boolean definition of pseudo-randomness, which is what allowed them to make that connection. In contrast, we say that a meta-polynomial computed by some circuit $C_n$ distinguishes one polynomial $f_n$ from another polynomial $g_n$ if $C_n$ outputs 0 when given the coefficient vector of $f_n$ as input, and outputs a nonzero value when given the coefficient vector of $g_n$. (By using interpolation, we can replace ``coefficient vector'' with ``vector of evaluations at sufficiently many points'' for any class which supports interpolation, that is, which is closed under affine linear transformations of the variables.) By only considering a meta-polynomial to distinguish one polynomial from another by its vanishing/non-vanishing, rather than in the probabilistic sense of pseudo-randomness, we are able to work over arbitrary fields, and make the connection with PIT instead of pseudo-random functions. 

We note, however, that if in their work one instead considers algebraic Turing machines (a la Blum--Shub--Smale) to distinguish functions---as they suggest at one point---then probability goes away. By considering the possible paths through such a machine, one gets a condition which is a logical combination of conditions on the vanishing/non-vanishing of certain polynomials, rather than the vanishing/non-vanishing of a single polynomial. See Remark~\ref{rmk:vanishing} for more details.

\section{Preliminaries}
A \emph{family} of polynomials $f=(f_n)$ consists of one polynomial $f_n$ for each $n$, usually on a number of variables that depends on $n$. A sequence of integers $a_1, a_2, \dotsc$ is \emph{p-bounded} if there is a polynomial $n^c + c$ such that $a_n \leq n^c + c$ for all sufficiently large $n$. A \emph{p-family} is a family of polynomials $(f_n)$ such that the number of variables of $f_n$ and the degree of $f_n$ are both p-bounded. We will primarily be interested in p-families throughout. 

A non-uniform algebraic complexity class is a collection of families of polynomials. $\cc{VP}$ is the collection of p-families $(f_n)$ such that $f_n$ computable by an algebraic circuit of $\poly(n)$ size. $\cc{VNP}$ is the collection of p-families $(g_n)$ such that there is a family $(f_n(x,e)) \in \cc{VP}$ such that $g_n(x) = \sum_{e \in \{0,1\}^{\poly(n)}} f_n(x,e)$. $\cc{VP}_{ws}$ is the collection of p-families $f=(f_n)$ such that $f_n(x) = \det_{\poly(n)}(L_n(x))$ where $L_n(x)$ is matrix whose entires are affine linear functions of the $x_i$. $\cc{\Sigma \Pi \Sigma}$ is the collection of p-families computable by polynomial-size, depth-three, layered circuits, with a linear combination gate at the output, preceded by a layer of multiplication gates, preceded by a layer of linear combinations of the input; that is, the polynomial is a sum of polynomially many products of linear functions of the inputs.

A polynomial $f(x_1, \dotsc, x_n)$ is a \emph{projection} of a polynomial $g(y_1, \dotsc, y_m)$ if there are affine linear functions $\ell_1(\vec{x}), \dotsc, \ell_m(\vec{x})$ such that $f(\vec{x}) = g(\ell_1(\vec{x}), \dotsc, \ell_m(\vec{x}))$, identically as polynomials. A polynomial family $f=(f_n)$ is a \emph{p-projection} of a polynomial family $g=(g_n)$ if there is a polynomial $t(n)$ such that for all $n$, $f_n$ is a projection of $g_{t(n)}$.

Let $\Poly^d(v)$ denote the space of \emph{homogeneous} polynomials of degree $d$ in $v$ variables, and $\Poly^{\leq d}(v)$ denote the space of (not necessarily homogeneous) polynomials of degree at most $d$ in $v$ variables. Homogeneity is used for technical simplicity; essentially everything we say can be modified to several other natural settings, such as non-homogeneous polynomials or multilinear polynomials.

Rather than the definitional viewpoint of a complexity class as a collection of families of polynomials, it will be useful to ``reverse the order of quantifiers'', and to consider, for each $n$, a subset of $\Poly^{d(n)}(v(n))$, and to consider a complexity class as a family of such subsets, one for each $n$. This viewpoint is implicit in much work on lower bounds in algebraic complexity theory, going back to work of Strassen (\eg, \cite{strassen}), and is explicit in geometric complexity theory (\eg, \cite{gct1, gct2, gctJACM, gct5, BLMW}). An example will help make this clear: In terms of lower bounds showing that some polynomial is not in $\cc{VP}_{ws}$, it is useful to think of $\cc{VP}_{ws}$ as being ``captured'' by the following family of sets:
\[
\mathcal{D}_n \defeq \{ f(x) \in \Poly^{\leq n}(n^2) : (\exists L)[f(x) = \det_n(L(x))]\}
\]
where the $L$ we consider here are those such that $L(X)$ is an $n \times n$ matrix whose entries are affine linear combinations of the variables $x_i$. The family $\mathcal{D}=(\mathcal{D}_n)_{n=1,2,3,\dotsc}$ captures $\cc{VP}_{ws}$ in the sense that, given a family of polynomials $g=(g_n)$, showing that $g_n \notin \mathcal{D}_{m_n}$ for all polynomially bounded sequences $(m_n)$ and for infinitely many $n$ proves that $g \notin \cc{VP}_{ws}$. We crystalize this into the following definition:

\begin{definition}
A family of subsets $(\mathcal{F}_n)$ with $\mathcal{F}_n \subseteq \Poly^{d(n)}(v(n))$ \emph{captures} a non-uniform algebraic complexity class $\mathcal{C}$ if:
\begin{enumerate}
\item For every family of polynomials $(f_n)$ with $f_n \in \mathcal{F}_n$ for all $n$, it follows that $(f_n) \in \mathcal{C}$, and

\item For every family of polynomials $f=(f_n) \in \mathcal{C}$, there is a polynomially bounded sequence of integers $m_n$ such that $f_n \in \mathcal{F}_{m_n}$ for every $n$. 
\end{enumerate}

We say that $(\mathcal{F}_n)$ \emph{captures $\mathcal{C}$ with padding} if we replace the last item by
\begin{enumerate}
\item[2$^\prime$.] For every family of polynomials $f=(f_n) \in \mathcal{C}$, there are polynomially bounded sequences of integers $e_n, m_n$ and a family of linear forms $\ell=(\ell_n)_{n=1}^{\infty}$ such that $\ell_n(x)^{e_n} f_n(x) \in \mathcal{F}_{m_n}$ for every $n$. \end{enumerate}

\end{definition}

It is readily seen that the family $\mathcal{D}_n$ above captures $\cc{VP}_{ws}$. Its homogeneous version,
\[
\mathcal{D}^h_n \defeq \{ f(x) \in \Poly^{n}(n^2) : (\exists L)[f(x) = \det_n(L(x))]\}
\]
where we only conisder \emph{linear} $L$ (zero constant term), captures homogeneous polynomials in $\cc{VP}_{ws}$ with padding.

Similarly, the family
\[
\mathcal{SPS}_n \defeq \left\{ f \in \Poly^{\leq n}(n) : \left[\exists a_{ijk} \in \F\right]\left(f = \sum_{i=1}^n \prod_{j=1}^{d(i)} \sum_{k=1}^n a_{ijk} x_k\right) \right\}
\]
captures $\cc{\Sigma \Pi \Sigma}$. While essentially all non-uniform algebraic complexity classes that are ever considered have a natural family of sets that captures them, note that such families are not unique. For example, $\cc{VP}_{ws}$ is also captured by the family of sets
\[
\mathcal{W}_n \defeq \{ f \in \Poly^{\leq n}(n) : f \text{ can be computed by a weakly-skew circuit of size } \leq n \}.
\]
While each $\mathcal{W}_n$ is quite different from each $\mathcal{D}_n$, this merely reflects the fact that weakly-skew circuit size and determinantal complexity are not equal, despite the fact that they are polynomially related.

Throughout, whenever we refer to a complexity class such as $\cc{VP}_{ws}(n)$, we really mean ``$\mathcal{F}_n$, for any fixed family $\mathcal{F}_n$ that captures $\cc{VP}_{ws}$ (possibly with padding).''

\subsection{Meta-polynomials and meta-complexity classes}
Given a space of polynomials $\Poly^{d_n}(v_n)$, we may consider \emph{meta-polynomials} on this space, which are polynomials $T$ whose variables correspond to the \emph{coefficients} of polynomials in $\Poly^{d_n}(v_n)$. That is, $T$ is a polynomial in $N = \binom{d_n + v_n-1}{d_n}$ variables. We denote the space of homogeneous meta-polynomials of degree $D$ by $\Poly^D(\Poly^{d_n}(v_n)) \cong \Poly^D(N)$. 
Given a polynomial $f \in \Poly^{d_n}(v_n)$ and a meta-polynomial $T \in \Poly(\Poly^{d_n}(v_n))$, 
we denote by $T(\coeff{f})$ the evaluation of $T$ at the coefficient vector of $f$. We will generally use capital letters to denote meta-polynomials, their degrees, and their number of variables, and lower-case letters for (non-meta) polynomials. 

\begin{example*}
The familiar polynomial $b^2 - 4ac$ may be considered as a meta-polynomial on the space $\Poly^2(2)$ of degree 2 homogeneous polynomials in 2 variables, namely, $\Poly^2(2) = \{ a x^2 + b xy + c y^2 : a,b,c \in \F\}$.  Then $T=b^2 - 4ac \in \Poly^2(\Poly^2(2))$, and evaluating $T$ at a polynomial $f = a x^2 + b xy + c y^2$ has the usual and natural meaning.
\end{example*}

We will want to consider families of meta-polynomials $T=(T_n)$ with $T_n \in \Poly(\Poly^{d_n}(v_n))$. If $v_n, d_n$ are themselves at least linear in $n$, then the number of variables of $T_n$ is exponential in $n$, so this family does not technically fit into the usual algebraic complexity classes as defined above. We would nonetheless like an analogue of the above classes where $T_n$ may depend on more than $\poly(n)$ variables, but its other relevant quantities are polynomial in its (usually much larger than $\poly(n)$) number of variables. We annotate such classes with a capital $N$, where $N_n$ is the number of variables of $T_n$ (in this case, $\dim \Poly^{d_n}(v_n)$). 

\begin{definition}[Stretched complexity classes] \label{def:stretched}
Given an algebraic complexity class $\mathcal{C}$, and a function $N(n)$, we define \emph{$\mathcal{C}$ with stretch $N$}, denoted $\mathcal{C}(N)$, as the class of families of polynomials $T = (T_n)$ such that there is a family $\overline{T} \in \mathcal{C}$ with $T_n = \overline{T}_{N(n)}$. 
\end{definition}

For most standard algebraic complexity classes, such as $\cc{VP}$, $\cc{VP}_{ws}$, $\cc{VNP}$, or $\cc{\Sigma\Pi\Sigma}$, this is equivalent to:

\begin{defStretchedPrime}[Alternative definition of stretched, for standard classes]
Given an algebraic complexity class $\mathcal{C}$, and a function $N(n)$, we define \emph{$\mathcal{C}$ with stretch $N$}, denoted $\mathcal{C}(N)$, as the class of families of polynomials $T = (T_n)$ such that $T$ satisfies the hypotheses of $\mathcal{C}$ with ``polynomial in $n$'' everywhere replaced by ``polynomial in $N(n)$.''
\end{defStretchedPrime}

To see that the two are equivalent: Given $T \in \mathcal{C}(N)$ according to Definition~\ref{def:stretched}$^\prime$, if we let $\overline{n}(N)$ be the inverse of $N(n)$ rounded to the nearest integer, then defining a family $\overline{T}_n = T_{\overline{n}(N(n))}$ satisfies Definition~\ref{def:stretched}. The opposite direction is clear.

For example, $\cc{VP}(N)$ denotes the class of families of polynomials $T=(T_n)$ where $T_n$ has $\poly(N)$ many variables, is of $\poly(N)$ degree, and can be computed by circuits of $\poly(N)$ size. We define $\cc{VNP}(N)$, $\cc{VP}_{ws}(N)$ and $\cc{\Sigma\Pi\Sigma}(N)$ analogously. 

Since we will typically be considering polynomials in $\poly(n)$ many variables with $\poly(n)$ degree, the space $\Poly^{d_n}(v_n)$ will have dimension $N_n = 2^{n^{O(1)}}$, so we have that $n = \poly(\log N)$. 

\section{An algebraic natural proofs barrier via polynomial identity testing}
We start by giving our definition of ``algebraic natural property;'' an algebraic natural proof in our sense will essentially be one that uses such a property. As is the case with Razborov--Rudich natural proofs, the latter is not a precise, formal definition, but in practice this will cause us no difficulties, and in particular does not affect our results (which are precise and formal). In the algebraic setting, a property of polynomials is a collection of subsets $C_n \subseteq \Poly^{d_n}(v_n)$ for some (usually p-bounded) sequences $d_n, v_n$. (Recall that everything we say is easily adapted to other kinds of polynomials such as non-homogeneous or multilinear.)

\begin{definition}[Natural property]
A property of polynomials $C=(C_n)$ with $C_n \subseteq \Poly^{d_n}(v_n)$ is \emph{natural} if it contains a set $C_n^* \subseteq C_n$ for each $n$ satisfying the following two conditions: 

\begin{enumerate}
\item \emph{Largeness: } $C_n^*$ is the complement of the zero-set of a meta-polynomial $T_n$. 

\item \emph{Constructivity: } The meta-polynomial family $T=(T_n)$ has degree and circuit size bounded by a polynomial in the number of its variables ($=\poly(\dim \Poly^{d_n}(v_n)) = \poly(\binom{d_n + v_n - 1}{d_n})$). That is, $T \in \cc{VP}(N)$ for $N_n = \dim \Poly^{d_n}(v_n)$.

\item \emph{Usefulness: } The algebraic circuit size of any family of functions $(f_n)$ with $f_n \in C_n$ for all $n$ is super-polynomial, that is, for any constant $d$, for sufficiently large $n$ the circuit size of $f_n$ is greater than $n^d$.
\end{enumerate}
\end{definition}

\begin{remark}[Deciding by (non)vanishing] \label{rmk:vanishing}
It is important for the connection with PIT that constructivity here be in terms of computing $T$ symbolically as a polynomial (or at least, some $T'$ such that $\{f : T'(f) \neq 0\} \subseteq C_n$), and not merely in terms of \emph{deciding} whether a given function $f$ is contained in $C_n^*$ (as is the case with Razborov--Rudich natural proofs). 

However, we note that even if we had allowed instead, say, Blum--Shub--Smale-style algebraic Turing machines to decide, given $\coeff{f}$,  whether or not $f \in C_n^*$, then much of the machinery still survives. In particular, the generic path through a BSS machine is still Zariski-open, being the intersection of finitely many Zariski-open subsets, and the ``yes/no'' output of the machine on generic inputs depends only on the vanishing/non-vanishing of a given polynomial. However, we would then need our hitting set to hit not only this final ``decider'' polynomial, but also all of the ``branching'' polynomials encountered along the generic computation path. If we wanted to consider all paths through the BSS machine, and not just the generic one, the situation becomes significantly more complicated, and as far as we are aware hitting sets for such computations have not been considered in the literature.
\end{remark}

\begin{remark}[Choice of field]
In terms of which fields to work over, in order to make the connection with derandomization, we want to work over fields that are large enough that derandomizing PIT over those fields is at least plausible. For simplicity, it may be easier to think of $\F$ as any infinite field. In principle, one could also work over a family of fields $\F_{s(n)}$ of size $s(n)$ greater than twice the degree of the polynomials under consideration (so the Schwarz--Zippell-DeMillo--Lipton Lemma holds). Note that for $s(n) < 2^{\poly(n)}$, the algebraic natural proofs barrier of Aaronson and Drucker also applies \cite{AD, ADblog}.
\end{remark}

We generalize this to:

\begin{definition}[$\Gamma$-natural against $\Lambda$]
For two complexity classes $\Gamma, \Lambda$, a property $C = (C_n)$ is \emph{$\Gamma$-natural against $\Lambda$} if it contains a subset $C_n^* \subseteq C_n$ satisfying:

\begin{enumerate}
\item \emph{Largeness: } $C_n^*$ is the complement of the zero-set of a meta-polynomial $T_n$. 

\item \emph{$\Gamma$-Constructivity: } The meta-polynomial family $T=(T_n)$ is in the meta-complexity class $\Gamma(N)$, where $N_n = \dim \Poly^{d_n}(v_n)$. 

\item \emph{Usefulness against $\Lambda$: } Any family of functions $f=(f_n)$ with $f_n \in C_n$ for all $n$ is not contained in $\Lambda$.
\end{enumerate}
\end{definition}

As observed in \cite{GrochowGCTUnity}, essentially all known algebraic circuit lower bounds to date are natural in this sense; in fact, most of them are $\cc{VP}_{ws}$-natural against the relevant complexity class, as they are defined by the rank of a matrix of size $\poly(N) \times \poly(N)$ (see Section~\ref{sec:idea}).

The key observation is the following. If there is a hitting set against $\cc{VP}$ which consists of the coefficient vectors of polynomials of number of variables, degree, and size $\poly(\log n)$, then there is no property that is $\cc{VP}$-natural against $\cc{VP}$. In other words, if for every meta-polynomial $T \in \cc{VP}(N)$, there is some polynomial $f \in \cc{VP}$ such that $T(\coeff{f}) \neq 0$, then one cannot prove a lower bound against $\cc{VP}$ by exhibiting a meta-polynomial that vanishes on a family of sets capturing $\cc{VP}$. As all such lower bounds to date are of this form \cite{GrochowGCTUnity}, and it is reasonable to expect future such lower bounds to be as well (see, \eg, \cite{strassen} or \cite[Appendix~B]{GrochowGCTUnity} for a more extended discussion of this expectation), this rules out quite a large class of lower bounds methods.

We formalize this observation with a definition and a theorem:

\begin{definition}[Succinct hitting set]
An algebraic complexity class $\Lambda$ is a \emph{succinct hitting set} against another class $\Gamma$ if there is a family of sets $\Lambda(n)$ which captures $\Lambda$, such that $\{\coeff{f} : f \in \Lambda(n)\}$ is a hitting set against $\Gamma(N)$, where $N$ is the dimension of the ambient space of $\Lambda(n)$. Namely, for all nonzero $T \in \Gamma(N)$, there is some $f \in \Lambda(n)$ such that $T(\coeff{f}) \neq 0$.
\end{definition}

\begin{theorem}
For any two algebraic complexity classes $\Gamma, \Lambda$, there is a $\Lambda$-succinct hitting set against $\Gamma$ if and only if there is no property which is $\Gamma$-natural against $\Lambda$.
\end{theorem}

We have essentially already given the proof in the paragraph above. \qed

The main open question is thus:

\begin{open}
Is $\cc{VP}$ a succinct hitting set against $\cc{VP}$? Is $\cc{VP}_{ws}$ a succinct hitting set against $\cc{VP}_{ws}$?
\end{open}

We note that it is not even obvious whether or not $\cc{VNP}$ is a succinct hitting set against $\cc{VP}_{ws}$. An important first step would be to show that known hitting sets against subclasses $\Gamma \subseteq \cc{VP}$ can be made $\Lambda$-succinct for as small a class $\Lambda$ as possible. For several pairs $(\Lambda, \Gamma)$ this is achieved in \cite{SV}. 

\begin{remark}[Generators]
A \emph{generator} for a class $\Gamma$ is a vector-valued function $\vec{G}(x_1, \dotsc, x_s)$ such that for any nonzero $f \in \Gamma$, $f(\vec{G}(\vec{x}))$ is not identically zero as a polynomial in $\vec{x}$. In other words, the image of $\vec{G}$---essentially an $s$-dimensional variety---is a hitting set (perhaps infinite) against $\Gamma$. The number of variables, $s$, is called the seed length of the generator; generators of small seed length are useful because they reduce PIT for $\Gamma$ from a many-variable problem to $s$-variable PIT, which is easily solved for small $s$. For most standard classes $\Lambda$, we note that if $\Lambda$ is a succinct hitting set against $\Gamma$, then this set is a generator against $\Gamma$ of small seed length. For most classes---such as $\cc{VP}, \cc{VNP}, \cc{VP}_{ws}$, $\cc{\Sigma \Pi \Sigma}$---are the image of a simply specified polynomial map $\vec{G}$ on few parameters. For example, the set of linear projections of the $n \times n$ determinant captures $\cc{VP}_{ws}(n)$ (with padding). This means that we may consider $\cc{VP}_{ws}(n)$ as the image of the map $M_{n^2 \times n^2} \to \Poly^n(n^2)$ which sends an $n^2 \times n^2$ matrix $L$ to the function $\det_n(L(\vec{x}))$, where we think of the $n \times n$ matrix $x$ simply as a vector of length $n^2$. If $\cc{VP}_{ws}$ is a hitting set for some class $\Gamma(N)$, then we may view it as a generator for $\Gamma(N)$ using the preceding encoding. The seed length of this generator is $n^4 = \poly(n)$ variables, but it outputs vectors in $\Poly^n(n^2)$, which has dimension $N$ that is exponential in $n$. So when $\cc{VP}_{ws}$ is a hitting set against some $\Gamma$, this generator still reduces from finding a hitting set in $N=2^{\Theta(n \log n)}$ variables to finding a hitting set in $n^4 = \poly(\log N)$ variables. As in the preceding example of $\cc{VP}_{ws}$ and the determinant, generators of small seed length are obvious for many classes; for $\cc{VP}$ this is somewhat less obvious, but is still true \cite{razElusive}.
\end{remark}

\section{Relationship with other topics in complexity}
\subsection{Geometric complexity theory} \label{sec:GCT}
In geometric complexity theory (GCT), the suggestion is not merely to use the polynomial method to find a meta-polynomial $T$ that vanishes on $\mathcal{C}_{easy}$ but not on some $f_{hard}$, but to additionally take advantage of the fact that most standard non-uniform classes $\mathcal{C}(n)$ are invariant under the action of some nontrivial group $G$, such as $\GL_n$ or $S_n$. This is because most measures of complexity do not depend on how we name the variables (leading to $S_n$ symmetry), and in many cases only change polynomially given a linear change of variables (leading to $\GL_n$ symmetry). The suggestion, without loss of generality, is thus to use a property $C_n$ to separate $\mathcal{C}(n)$ from $f_{hard}$ such that $C_n$ is also sent to itself by the same symmetry group. In this case, rather than considering a single meta-polynomial $T$, we may, again without loss of generality, consider the entire linear span $V$ of all meta-polynomials $T'$ that are in the $G$-orbit of $T$. (When $G$ is $S_n$ it is clear that $V$ is finite-dimensional; even over infinite fields, however, this is also true of the $\GL_n$-orbit of $T$.) $V$ is then a representation of $G$ or $G$-module; following \cite{GrochowGCTUnity} we refer to a $G$-module of meta-polynomials as a ``test $G$-module,'' since its vanishing is a test for having a given $G$-invariant property.

For $G=\GL_n$ (a natural group of symmetries for many standard algebraic circuit classes such as $\cc{VP}$, $\cc{VP}_{ws}$, $\cc{VNC}$, $\cc{VNP}$, $\cc{VQP}$, $\cc{\Sigma \Pi \Sigma}$), every irreducible $G$-module contains an essentially unique highest weight vector (see, \eg, \cite{fultonHarris}) (=highest weight test polynomial), which is an ``HWV obstruction'' in the terminology of \cite{BI}. (Conversely, every HWV obstruction gives rise to a test module.) Considering these HWV obstructions directly, \Burgisser and Ikenmeyer were able to prove lower bounds on matrix multiplication using the technology of GCT \cite{BI}. This raises the natural question of: given the label $\lambda$ of an irreducible $\GL_n$-module ($\lambda$ is a partition with at most $n$ parts, see, \eg, \cite{fultonHarris}), how computationally hard is it to construct its (unique) highest weight test polynomial? However, from the viewpoint of algebraic natural proofs, we are led to a related but slightly different question.

The first natural question to think of is to determine the circuit complexity of the HWV obstructions. However, algebraic natural proofs suggests asking something still further.

Namely, suppose that $\Gamma$-natural proofs cannot prove lower bounds against $\Lambda$, and suppose that $\Lambda(n)$ is invariant under a group $G_n$ (not necessarily $\GL_n$---in particular, we do not need the theory of highest weights for what we are about to say). Then given a sequence of test $G_n$-modules $V_n$, potentially useful against $\Lambda(n)$, if there is a sequence of meta-polynomials $T_n \in V_n$ such that $(T_n)_{n=1,2,3,\dotsc}$ is in $\Gamma$, then for infinitely many $n$, $V_n$ is not useful against $\Lambda(n)$---that is, $V_n$ does not vanish identically on $\Lambda(n)$. We are thus led to the question:

\begin{open} \label{q:complexityTestModule}
For any given sequence of test $G$-modules $V_n$, what is the complexity of the \emph{easiest} family of meta-polynomials $(T_n \in V_n)$?
\end{open}

In particular, while the complexity of any given $T_n \in V_n$ doesn't change within the orbit of $T_n$, $V_n$ itself contains all \emph{linear combinations} of points on this orbit, and some such linear combinations could have significantly lower complexity than, say, the HWVs in $V_n$ (when $V_n$ is a test $\GL_n$-module).

Note that, whether or not there is a natural proofs barrier for $\cc{VP}$, the above question is interesting. For if there is such a barrier, then any family of test $G$-modules with low-complexity polynomials cannot be used to prove lower bounds.\footnote{It is interesting to note that if, for a given sequence of labels $\lambda(n)$, we could find an upper bound on the easiest family of test polynomials in \emph{any} family of test $\GL_n$-modules isomorphic to $V_{\lambda(n)}$, then this could be used to rule out \emph{multiplicity} obstructions. At the moment, there are essentially no techniques known for ruling out multiplicity obstructions, only for ruling out occurrence obstructions, e.\,g., \cite{ikenmeyerPanova,BIP,GIP}.}
Conversely, if there is no such barrier, then any family of test $G$-modules with low-complexity polynomials might be a good place to look for test polynomials to prove lower bounds, since we might hope that low-complexity test polynomials would be easier to understand and therefore easier to use to try to prove lower bounds.

This question is perhaps more immediately interesting in the following specific cases: Given a class $\Gamma$ for which it is shown in \cite{SV} that $\Gamma$-natural proofs cannot prove lower bounds against $\cc{VP}$, which families $V_{\lambda(n)}$ of test $\GL_n$-modules contain a family of test polynomials $T_n$ such that $(T_n) \in \Gamma$? Note that, even for test $\GL_n$-modules $V$, the highest weight meta-polynomials need not be the easiest polynomials in $T$.  So although considering HWV obstructions may be useful for proving lower bounds, in order to prove that certain test $\GL_n$-modules are \emph{not} useful for lower bounds, one needs to consider the more general Open Question~\ref{q:complexityTestModule}.

\subsection{Algebraic proof complexity}
Pitassi \cite{pitassi96, pitassiICM} and Grochow \& Pitassi \cite{GP} introduced the Ideal Proof System (IPS), for refuting unsatisfiable CNFs using algebraic reasoning. While IPS is a very strong proof system---at least as strong as Extended Frege---they also introduced a variant of this system called the \emph{Geometric} IPS (it is an open question whether Geometric IPS can p-simulate general IPS). Using the connection in this paper it may be plausible to prove unconditional lower bounds against Geometric IPS.  We now discuss this in a bit more detail.

\begin{definition*}[{Geometric Ideal Proof System, ``Geometric IPS,'' \cite[Appendix~B]{GP}}]
Given an unsatisfiable system of polynomial equations $f_1(\vec{x}) = \dotsb = f_m(\vec{x}) = 0$, a \emph{geometric IPS certificate} of unsatisfiability consists of an algebraic circuit $C(y_1, \dotsc, y_m)$ such that
\begin{enumerate}
\item $C(\vec{0}) = 1$, and

\item $C(f_1(\vec{x}), \dotsc, f_m(\vec{x})) = 0$, in other words, $C$ is a polynomial relation amongst the $f_i$.
\end{enumerate}
For any algebraic circuit class $\mathcal{C}$, a \emph{geometric $\mathcal{C}$-IPS proof} is an algebraic circuit in $\mathcal{C}$ on inputs $y_1, \dotsc, y_m$ computing some geometric IPS certificate.
\end{definition*}

This system may be used to prove that a 3CNF formula is unsatisfiable as follows. Given a 3CNF formula with $m$ clauses, we translate it into a system of $m$ polynomials of degree at most $3$ in the natural way, so that any Boolean assignment to the variables satisfies a clause iff the corresponding polynomial evaluates to 0. Then the 3CNF is unsatisfiable iff the corresponding equations $f_1(\vec{x}) = \dotsb = f_m(\vec{x}) = x_1^2 - x_1 = \dotsb = x_n^2 - x_n = 0$ are unsatisfiable over $\F$. In \cite[Appendix~B]{GP} it shown that geometric IPS, without any complexity bounds on the circuit computing a certificate, is a sound and complete proof system for such systems of equations. In fact, over any algebraically closed field or any dense subfield of $\mathbb{C}$, they showed that the same is true even if the equations $x_i^2 - x_i = 0$ are omitted. 

The idea of the geometric IPS is to consider the equations $f_1(x_1, \dotsc, x_n), \dotsc, f_m(x_1, \dotsc, x_n)$ as a map $f\colon \F^n \to \F^m$, and to note that the system of equations $f_1 = \dotsc = f_m = 0$ is satisfiable iff 0 is in the image of the map $f$. A geometric IPS certificate proves that, not only is 0 not in the image, but 0 is not even in the \emph{closure} of the image of the map $f$. The geometric object of interest here is thus the image of the map $f$.

Suppose we have a family $(\mathcal{F}_n)_{n=1,2,\dotsc}$ of systems of polynomial equations 
\[
\mathcal{F}_n = (f_{n,1}(x_1, \dotsc, x_{n^c}), \dotsc, f_{n,n^d}(\vec{x})),
\]
 such that the images of the maps $f_{n}\colon \F^{n^c} \to \F^{n^d}$ are a hitting set against some circuit class $\Lambda$. Then, by condition (2) of the above definition, no geometric $\Lambda$-IPS certificate can exist. Although here we are using the evaluations of polynomials rather than their coefficient vectors, note that for any class $\Lambda$ capable of interpolation---that is, closed under affine linear transformations---a succinct hitting set can be defined either in terms of coefficient vectors or in terms of the vector of evaluations at sufficiently many points.
 
\begin{open}
For various $\Lambda$ for which hitting sets are known, prove lower bounds on the Geometric $\Lambda$-Ideal Proof System by finding a succinct hitting set of the following form: there is a family of unsatisfiable 3CNFs $(\varphi_n)$ such that, if $f_n$ is the above polynomial map associated to $\varphi_n$, then the image of $f_n$ is a hitting set against $\Lambda$. 
\end{open}
 
Of course, it would also be interesting to show that for certain $\Lambda$ no hitting sets of this form exist. 

Unfortunately, we were unable to get the same connection to work for general IPS. The natural object to look at for general IPS is not the image of $f$, but rather its graph $\{(\vec{\alpha}, \vec{f}(\vec{\alpha})) : \vec{\alpha} \in \F^n\}$. The issue is that, when the $f_i$ are themselves described by small circuits, as is essentially always the case in instances of complexity-theoretic interest, the function $y_i - f_i(\vec{x})$ is a very easily computable function which vanishes on the graph of $f$.

\begin{open}
Find and exploit an analogous connection between algebraic natural proofs / hitting sets and (general) IPS.
\end{open}
 
\section*{Acknowledgments}
We thank Scott Aaronson and Andy Drucker for conversations about their work \cite{AD, ADblog}, the relationship between the two approaches and how they might be combined. 
We thank Amir Shpilka for conversations related to Section~\ref{sec:GCT} and for his encouragement to publicize our thinking, even in light of the results of \cite{SV} which (independently) supercede ours. 
During the course of this work, J.A.G. was supported by A. Borodin's NSERC Grant \# 482671, an Omidyar Fellowship from the Santa Fe Institute, and NSF grant DMS-1620484; M.S. was supported by NSF grant CCF-1218711 and by Simons Foundation Award 332622; and S.S. was supported in part by NSF grant CCF-1350572.

\bibliographystyle{plainurl}
\bibliography{algnatpfs}

\end{document}